\documentclass[aip,reprint]{revtex4-1}

\usepackage{graphicx}
\usepackage{gensymb}
\usepackage{mathrsfs}
\begin{document}

\title{Single-Shot Electron Diffraction using a Cold Atom Electron Source}

\author{Rory W. Speirs}
\affiliation{School of Physics, The University of Melbourne, Victoria 3010, Australia}
\author{Corey T. Putkunz}
\affiliation{School of Physics, The University of Melbourne, Victoria 3010, Australia}
\author{Andrew J. McCulloch}
\affiliation{School of Physics, The University of Melbourne, Victoria 3010, Australia}
\author{Keith A. Nugent}
\affiliation{ARC Centre of Excellence for Advanced Molecular Imaging, Department of Physics, La Trobe University, Victoria 3086,
Australia}
\author{Benjamin M. Sparkes}
\affiliation{School of Physics, The University of Melbourne, Victoria 3010, Australia}
\author{Robert E. Scholten}
\email{scholten@unimelb.edu.au}
\affiliation{School of Physics, The University of Melbourne, Victoria 3010, Australia}

\date{\today}

\begin{abstract}
Cold atom electron sources are a promising alternative to traditional photocathode sources for use in ultrafast electron diffraction due to greatly reduced electron temperature at creation, and the potential for a corresponding increase in brightness. Here we demonstrate single-shot, nanosecond electron diffraction from monocrystalline gold using cold electron bunches generated in a cold atom electron source. The diffraction patterns have sufficient signal to allow registration of multiple single-shot images, generating an averaged image with significantly higher signal-to-noise ratio than obtained with unregistered averaging. Reflection high-energy electron diffraction (RHEED) was also demonstrated, showing that cold atom electron sources may be useful in resolving nanosecond dynamics of nanometre scale near-surface structures.
\end{abstract}


\maketitle 

\section{Introduction}
Ultrafast single-shot diffraction is revolutionising our understanding of materials science, chemistry, and biology, by imaging objects on atomic length and time scales simultaneously \cite{Barty2008, Dwyer2006, Zewail2006}. X-ray free electron lasers (XFELs) have been used to perform single-shot coherent diffractive imaging on micro- and nano-metre scale targets \cite{Seibert2011, Chapman2006}, where the imaging pulse is briefer than the time scale of damage to the object \cite{Neutze2000, Quiney2011}. An alternative and complementary approach is ultrafast electron diffraction, which benefits not only from much stronger scattering of electrons relative to X-rays \cite{Carbone2012}, but also significantly reduced damage per elastic scattering event \cite{Henderson1995}. To enable single-shot diffraction studies, the number of electrons per pulse must be of order $10^6$ or greater to have sufficient signal per pixel at the detector \cite{Krivanek1993}. This number is easily achievable with photocathode sources, and when combined with RF bunch compression, sub $100\,\rm fs$ pulses have been achieved \cite{vanOudheusden2010}. The brightness of photocathode sources is limited by the high initial temperature of the extracted electrons ($10^4\,\rm K$), leading to a high transverse emittance \cite{Claessens2005}. The emittance required for single-shot imaging depends on the size of the object being imaged: larger object sizes or crystal periods require lower emittance to generate useful coherent diffraction patterns. Ultrafast single-shot electron diffraction has been achieved from large single crystal and polycrystalline samples using a variety of photocathode based sources \cite{Tokita2009, Li2010, Musumeci2010, Sciaini2009, Siwick2003}, but insufficient source brightness has prevented demonstration for micro or nanocrystals, or single molecules.

Cold electron sources are a promising alternative to solid state photocathodes, producing electrons by near threshold photoionisation of laser cooled atoms. Electrons from these sources have an intrinsic temperature as low as $10\,\rm K$, which for a given flux leads to several orders of magnitude increase in brightness \cite{McCulloch2013, Engelen2013}. The low electron temperature, together with the ability to spatially shape the beam \cite{McCulloch2011}, should allow these sources to produce uniformly filled ellipsoid bunches, which do not suffer emittance degradation resulting from non-linear internal Coulomb forces \cite{Luiten2004}.

A cold electron source was recently used for the first time to generate a transmission electron diffraction pattern \cite{Mourik2014}. In that experiment, cold, sub-picosecond electron pulses containing a few hundred electrons were scattered by graphite. To produce diffraction patterns with clearly discernible Bragg reflections, several thousand individual shots were integrated.

Here we present the first \emph{single-shot} electron diffraction patterns obtained using a cold electron source. The patterns were obtained from a monocrystalline gold foil using a single $5\,\rm ns$ bunch of $5\times 10^5$ electrons. No electron aperture was required due to the high spatial coherence of the electrons at the source. This allowed all generated electrons to contribute to the image, resulting in a single shot with sufficiently high signal-to-noise ratio for identification of the sample and registration of successive images. Single-shot reflection high-energy electron diffraction (RHEED) was also demonstrated from a wafer of monocrystalline silicon.

\section{The Cold Atom Electron Source}
\begin{figure*}
	\centering
	\includegraphics[width=170mm]{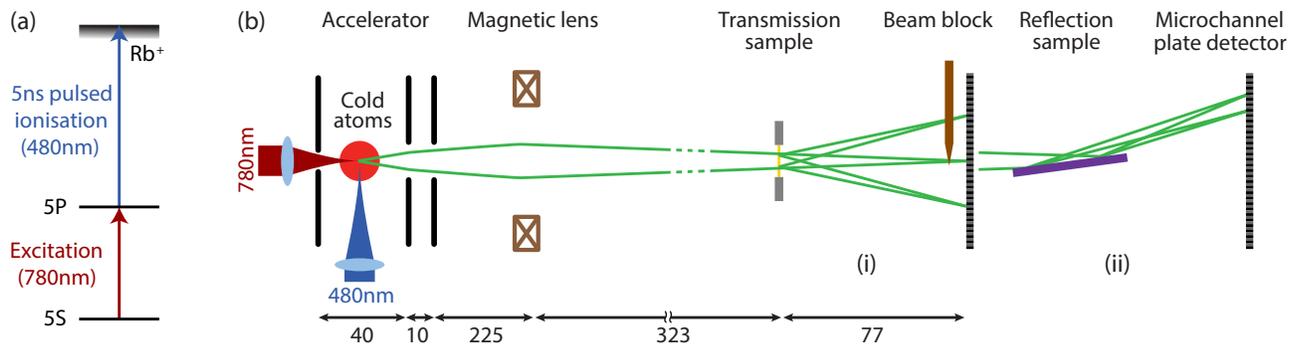}
	\caption{(a) Rubidium atoms are ionised in a two step process: $780\,\rm nm$ laser light drives them to the first excited state where they are ionised by a $5\,\rm ns$ pulse of $480\,\rm nm$ light. (b) The electrons produced are accelerated by a static electric field, focused, and scattered off a sample, either in transmission (i) or reflection (ii) geometries. Distances are in millimetres. \label{figure1}}
\end{figure*}

The cold atom electron source (CAES) generates electrons by photoionisation of rubidium-85 atoms in a magneto-optical trap, which is positioned between two accelerating electrodes as shown in figure \ref{figure1}b.

The photoionisation is a two-stage process (figure \ref{figure1}a). The atoms are excited from the $5S_{1/2} (F=3)$ ground state to the $5P_{3/2} (F=4)$ excited state using a $100\,\rm ns$ pulse of laser light of wavelength $780\,\rm nm$. A $5\,\rm ns$ pulse from the ionisation laser (wavelength $480\,\rm nm$) then drives the atoms either to a Rydberg level, or directly to the continuum. 

The excitation laser illuminates the atom cloud along the axis of electron propagation (longitudinal direction), and the focused intensity profile can be changed arbitrarily using a liquid crystal spatial light modulator, which defines the shape of the electron bunch in the two dimensions transverse to propagation \cite{McCulloch2011}. The blue ionisation laser illuminates the atom cloud transversely to the electron propagation axis, defining the longitudinal profile of the electron bunch which is generated in the region of overlap of the two laser beams such that the bunch is shaped in all three dimensions. 

The ionisation time is determined by the temporal profile of the blue tunable dye laser pulse, with full width half maximum (FWHM) duration of $5\,\rm ns$, pulse energy of $5\,\rm mJ$ at the cloud, and repetition rate of $10\, \rm Hz$. The blue laser is focused with a cylindrical lens onto the atom cloud, so that the ionisation region is defined by a ribbon of light with a FWHM width of $30\,\rm \mu m$ in the longitudinal direction. 

Before ionisation the atom cloud has a peak density of approximately $10^{10}\,\rm atoms\,cm^{-3}$, and temperature $100\,\rm \mu K$. The quadrupole magnetic field is switched off and allowed to decay for $3.5\,\rm ms$ before ionisation, however a small magnetic field is still present after this time which slightly alters the electron trajectory from shot-to-shot.

The accelerator can be used in a two or three electrode configuration. The arrangement of applied potentials allows flexibility in determining the extraction electric field strength, the final energy of the electrons, and the diverging beam angle, which is defined in part by the lensing effect of the electrodes. In addition, the energy spread of the extracted electrons is determined by the combination of extraction electric field strength and longitudinal width of the ionisation region. A standard configuration of potentials with field strength $2.6\,\rm kVcm^{-1}$ and a blue beam width of $30\,\rm \mu m$ results in an electron energy spread of $8\,\rm eV$. This is a relatively high energy spread compared to sources used in conventional electron microscopes, where chromatic aberration in the strong objective lens drives the need for low energy dispersion. The contribution to the point spread function due to chromatic aberration is proportional to the beam semi-angle accepted into the lens, and for the single weak condenser lens used in our setup, this contribution is significantly smaller than the detector resolution. Polychromaticity also results in a spread of diffraction angles for any given sample spatial frequency, limiting the achievable resolution in coherent diffractive imaging. We typically use an electron energy of $8\,\rm keV$ for diffraction experiments, giving a fractional energy spread of $\Delta E/E<0.001$, which contributes negligibly to the spread in diffraction angles from the sample, and would allow coherent diffractive imaging of objects $20\,\rm nm$ across to a resolution of better than $1\,\rm \AA$ \cite{Zurch2014}. The energy spread could be reduced further if required by tailoring the extraction field strength or reducing the focal spot size of the blue laser beam, though the latter method would also reduce the number of electrons generated.

After the electron bunch leaves the accelerator, it traverses a solenoid magnetic lens at a distance of $225\,\rm mm$, before drifting $323\,\rm mm$ to the sample. The low numerical aperture of the lens limits the ability to create very small beam sizes at the sample, but results in a highly collimated beam without the need to introduce further electron optics. After the target sample the diffracted electrons propagate $77\,\rm mm$ to a phosphor-coupled microchannel plate (MCP) detector which is imaged with a camera.

\section{Beam Parameters}
We used a Gaussian excitation laser beam with a FWHM width of $80\,\rm \mu m$ at the focus. However, as our intension was to ionise as many electrons as possible, we used a very high power in the beam (with a peak intensity of thousands of times the saturation intensity). This resulted in significant excitation even a long way from the centre of the beam, as well as significant fluorescence and reabsorption, which both have the effect of increasing the excitation area. The excited area was ultimately determined by measuring the unfocused electron bunch size at the detector, along with the known magnification of the beam path. Using this method, the electron bunch at the source was determined to be a peaked shape, with a FWHM width of $1.4\,\rm mm$. From previous measurements \cite{McCulloch2013}, electrons generated in this method are known to have a source divergence of $\sigma_{\theta_x}=0.3\,\rm \mu rad$, resulting in a source emittance for the bunch generated here of $\epsilon_x=50\,\rm nm\,rad$. The electrons were focused to a minimum spot size at the MCP as shown in figure \ref{figure1}, resulting in a beam width at the sample of approximately $300\,\rm \mu m$, with a corresponding coherence length at this point of $\ell_c=2\,\rm nm$.

Using a Faraday cup, the number of electrons per pulse was measured to be $5\times10^5$ ($80\,\rm fC$), corresponding to an ionisation fraction of approximately $50\%$ within the ionisation region of the atom cloud when taking into account the density of the cloud, and the volume of the illuminated region. 

The bunch temporal profile was estimated to be Gaussian with a FWHM duration of $5\,\rm ns$ based on the pulse length of the blue laser. While the actual electron profile may differ from this significantly due to effects like a rapid depletion of excited state atom population, or an extended laser tail, $5\,\rm ns$ is expected to offer a fairly good characteristic timescale over which the bunch is produced. Combining the measured electron number per pulse, and the estimated temporal profile, yields a peak beam current of $20\,\rm \mu A$. When combined with the estimated emittance, this gives a peak brightness of $\mathscr{B}=3\times10^{8}\,\rm A\,m^{-2}sr^{-1}$, using the same conventions as in ref. \cite{McCulloch2013}.

The emittance and bunch charge for the CAES are therefore approaching the values required for single-shot diffraction of microcrystals \cite{vanOudheusden2007}, but the pulse duration is possibly still three orders of magnitude too long to avoid degradation of the diffraction pattern due to beam induced damage of such small samples. However recent studies have suggested the constraints on pulse duration due to damage could be relaxed for electrons compared with X-rays, because of the differences between the scattering and damage-inducing processes \cite{Egerton2015}. To achieve sub-picosecond ultrafast electron diffraction, the ionisation process can be modified to use femtosecond rather than nanosecond laser pulses \cite{McCulloch2013, Engelen2013}. Picosecond to femtosecond duration electron bunches containing the same charge will require spatial beam shaping in order to avoid brightness degradation caused by the otherwise non-linear space charge expansion of the bunch \cite{Luiten2007}.

\section{Single-Shot Electron Diffraction}
\begin{figure}
	\centering
	\includegraphics[width=85mm]{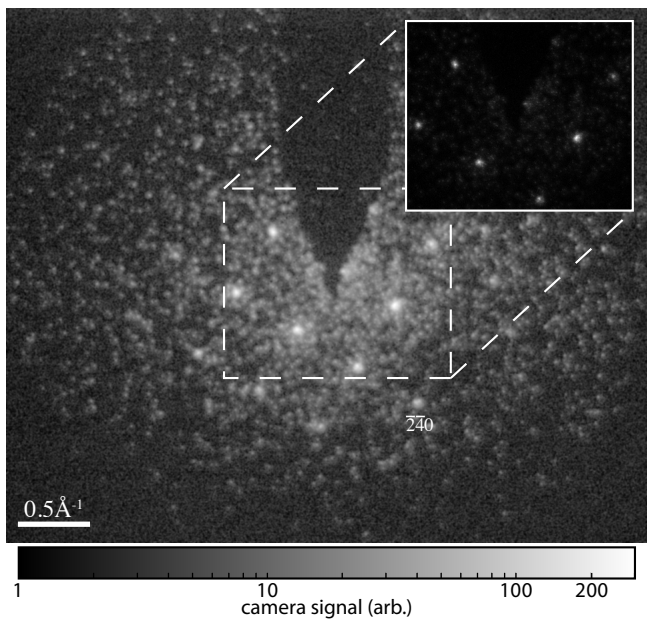}
	\caption{Single-shot transmission electron diffraction from gold, formed from a $5\,\rm ns$ pulse of cold electrons. Main image is logarithmically scaled, inset is linearly scaled.\label{figure2}}
\end{figure}

We demonstrated diffraction of electron bunches generated in the CAES from an $11\,\rm nm$ thick gold foil mounted on a $3\,\rm mm$ transmission electron microscopy grid, with an electron energy of $8\,\rm keV$. Figure \ref{figure2} shows the diffraction pattern from a single $5\,\rm ns$ electron bunch. The resulting Bragg reflections are clearly visible out to the $(\overline{2}\overline{4}0)$ spot at a resolution of $1.1\,\rm \AA^{-1}$, where the crystallographic convention has been adopted for reciprocal lattice vectors, such that $\left|\mathbf{g}_{hkl}\right| = 1/d_{hkl}$, where $d_{hkl}$ is distance between atomic planes in real space.

\begin{figure}
	\centering
	\includegraphics[width=85mm]{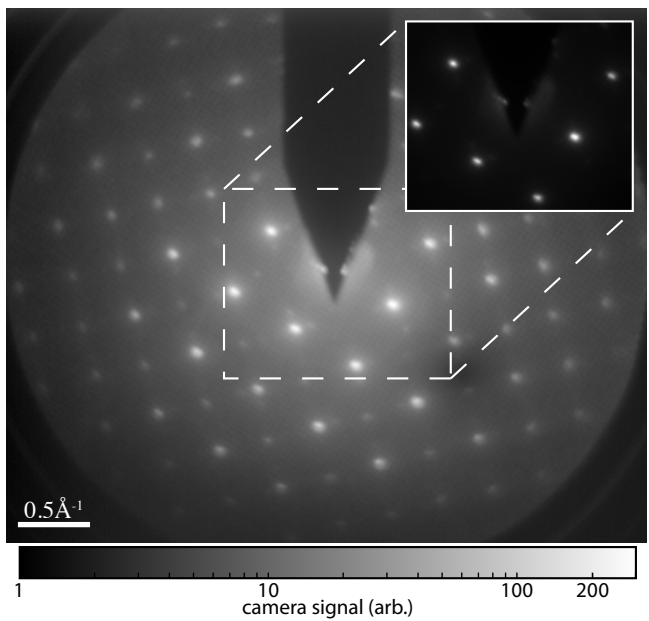}
	\caption{2000 diffraction shots from gold, directly averaged. Averaging results in a higher signal-to-noise ratio, but shot-to-shot beam instabilities lead to a broadening of the Bragg peaks. Main image is logarithmically scaled, inset is linearly scaled.\label{figure3}}
\end{figure}

The reflections show the four-fold symmetry of the gold face-centered cubic (fcc) lattice, and the $\{200\}$ and $\{220\}$ reflections visible on the sides and corners of the inner square confirm the unit cell parameter as $0.407\,\rm nm$, consistent with the agreed value for gold of $0.40782\,\rm nm$ at $25\rm \degree C$ \cite{Maeland1964}.
To obtain a higher signal-to-noise ratio, 2000 shots were averaged. The result (figure \ref{figure3}) allows Bragg spots to be identified out to the $(\overline{6}60)$ reflection, with an effective resolution of $2.08\,\rm \AA^{-1}$, limited by the size of the detector.

Closer inspection reveals that the Bragg spots have been significantly broadened when compared to the single-shot case, indicating that the transverse coherence of the time-averaged electron beam is reduced compared to any single constituent bunch. This loss of coherence stems from a slight beam wobble due to small variations in the decay of the quadrupole magnetic field after it is switched off. The effect of the beam drift on the diffraction pattern is analogous to the varying shot-to-shot diffraction patterns obtained in XFEL nanocrystal diffraction experiments, where diffraction patterns are obtained from millions of randomly aligned nanocrystals \cite{Chapman2011}.

\begin{figure}
	\centering
	\includegraphics[width=85mm]{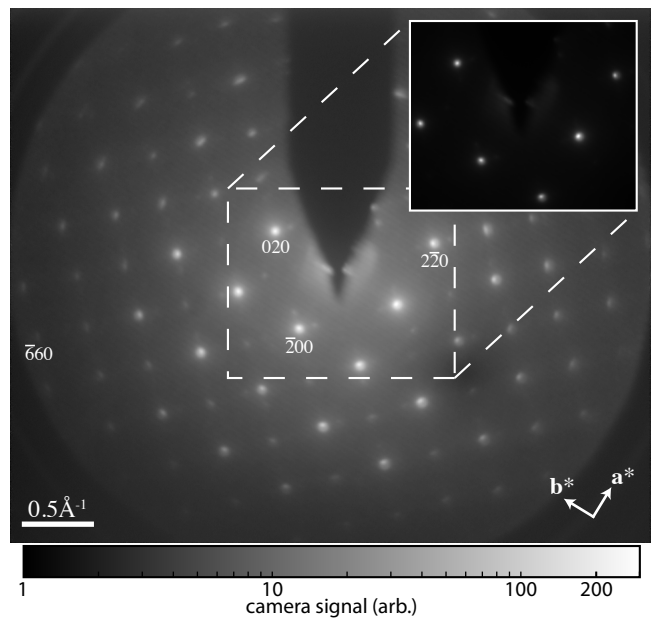}
	\caption{2000 diffraction shots from gold averaged by registering individual images. Registration is possible due to the high signal-to-noise ratio in the single shots, and recovers the sharpness in Bragg peaks lost in direct averaging. The reciprocal lattice vectors $\mathbf{a^*}$, $\mathbf{b^*}$ are drawn to scale. Main image is logarithmically scaled, inset is linearly scaled.\label{figure4}}
\end{figure}

To compensate for the beam wobble, successive single-shot images were registered. The eleven brightest spots were used to adjust the alignment by performing a cross correlation of the individual images and the unregistered average in the region surrounding the spots. The resulting registered average can be seen in figure \ref{figure4}, showing notably sharper Bragg spots.

\begin{figure}
	\centering
	\includegraphics[width=85mm]{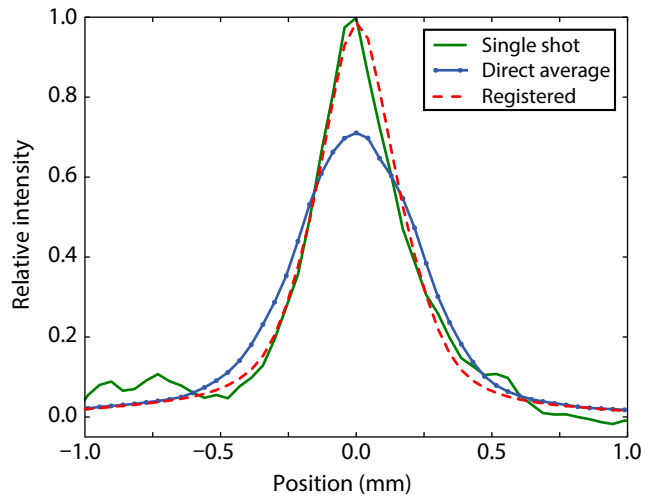}
	\caption{A lineout of the $({\overline{2}00})$ Bragg reflection in the $\mathbf{b^*}$ direction. Registering the single shots averages out the noise without resulting in Bragg spot broadening, as happens when shots are directly averaged. \label{figure5}}
\end{figure}

A lineout (figure \ref{figure5}) of the $({\overline{2}00})$ Bragg reflection in the $\mathbf{b^*}$ direction shows how the direct average reduces the noise level compared to a single shot, at the expense of increasing peak width. The registered average maintains the low noise level of the direct average, while fully recovering the peak resolution. These results emphasise that the imaging is indeed effectively single-shot, with features clearly visible above the noise out to a resolution to $1.1\,\rm \AA^{-1}$.

\begin{figure}
	\centering
	\includegraphics[width=85mm]{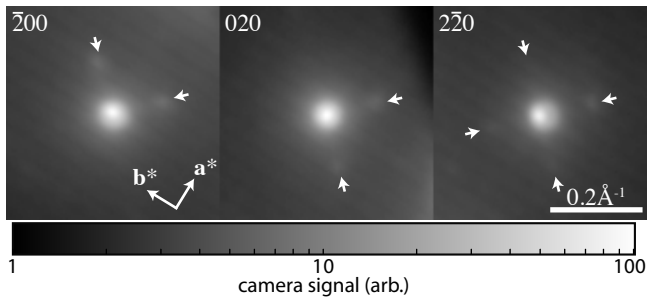}
	\caption{The two satellite spots around the $({\overline{2}00})$ and $(020)$ reflections, and the rightmost spot around the  $({2\overline{2}0})$ reflection, are due to diffraction from crystal twins. The other spots visible around the $({2\overline{2}0})$ reflection are due to the electron diffracting twice: once each from two different twins. The small arrows indicate the positions of the faint satellite spots. The reciprocal lattice vectors are not to scale.\label{figure6}}
\end{figure}

Due to the structure amplitude of fcc gold, the only allowed reflections are those where the Miller indices $h, k, l$, are all even or all odd. The diffraction images of gold were taken along the $\langle001\rangle$ zone axis, where the lattice amplitude dictates that reflections are only allowed if $h$ and $k$ are even. While at low diffraction angles these rules are obeyed, it can be seen in figures \ref{figure3} and \ref{figure4} that some kinematically disallowed reflections are visible at higher angles. This is caused by a departure from the single scattering kinematic approximation, where both elastically, and inelastically scattered electrons are then re-scattered into directions which are forbidden to the unscattered incident beam. It can also be seen from figure \ref{figure6} that the Bragg reflections are accompanied by two or more satellites, offset from the main reflection at $45\degree$ from the direction of the reciprocal basis vectors.

These satellites are the result of $\{111\}$ crystal twinning, which can form when $(100)$ gold films are prepared by evaporation \cite{Pashley1963}. There are two different mechanisms behind the satellite creation. The two relatively strong satellites that can be seen around the $({\overline{2}00})$ and $(020)$ Bragg reflections are created directly by diffraction from the crystal twins. This also produces the rightmost spot around the $({2\overline{2}0})$ reflection. The other satellites around the $({2\overline{2}0})$ reflection arise from double diffraction, where electrons are first diffracted by a $(100)$ oriented domain, and then diffracted again by one of the twins. Dynamical effects would not normally be seen with such a thin sample because electron energies used in a transmission electron microscope are typically ten times greater than used here, resulting in a much lower scattering cross section and interactions that more closely match the simple kinematic theory.

\begin{figure}
	\centering
	\includegraphics[width=85mm]{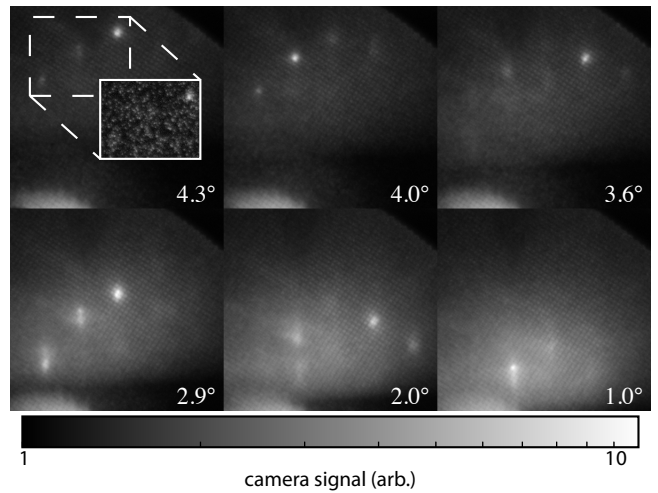}
	\caption{100 shot averages of RHEED from silicon $\langle100\rangle$ at a range of polar angles. Inset: a single shot of the region indicated, clearly showing a Bragg reflection.\label{figure7}}
\end{figure}

Reflection high-energy electron diffraction (RHEED) is a surface sensitive diffraction technique routinely used to monitor crystal surface quality and epitaxial crystal growth \cite{Norton1983}, and has also been used to observe surface dynamics resulting from illumination by ultrafast lasers \cite{Herman1990}. RHEED is a useful technique to further demonstrate diffraction from our source, both because the electron energies typically required fall into the range we can easily generate, and because high quality single crystals are more readily available as bulk wafers than as the nanometre-thick foils needed for transmission electron diffraction. To adjust the system for RHEED, all that was required was to rotate the sample through 90 degrees as shown in figure \ref{figure1}b(ii). Figure \ref{figure7} shows RHEED patterns from a $\langle100\rangle$ silicon wafer, which was HF etched to remove the native oxide layer immediately prior to transfer to the vacuum chamber. The beam was nominally incident on the crystal from the $[110]$ direction at $0\degree$ polar angle. Since the sample stage could not be rotated in the azimuthal direction, it is likely that the wafer was slightly misaligned, which would account for the apparent horizontal asymmetry of the Bragg reflections at any particular polar angle. For clarity the RHEED patterns shown are 100 shot averages, however Bragg reflections were easily visible from a single shot as shown by the inset.

Cold electron RHEED offers a promising opportunity to investigate near-surface dynamics on nanosecond time scales. The high transverse coherence of the beam should also allow coherent scattering to be observed from structures tens of nanometres wide, such as quantum dots and optical metamaterials. Cold electron sources with bunch shaping to control space-charge induced brightness degradation are perhaps uniquely placed to perform these studies, due to their potential to deliver high bunch charges and high coherence at relatively low electron energies. Using very high electron energies to mitigate space-charge effects is not an option for RHEED, since very energetic electrons penetrate too deeply to accurately probe surface structure.

\section{Summary}
We have demonstrated single-shot electron diffraction using fast electron bunches produced with a cold atom electron source. The $5\,\rm ns$ bunches contained around $5\times 10^5$ electrons, and because of their low temperature and high coherence, 
no beam aperture was required, allowing all generated electrons to contribute to imaging. When scattered by a single crystal of gold, the resulting single-shot diffraction pattern contained sufficient signal to give information about the crystal structure without averaging. The large signal-to-noise ratio allowed subsequent shots to be merged through image registration, which compensated for shot-to-shot beam drift that degraded the image quality when directly summing. Single-shot diffraction pattens have also been obtained in reflection mode, which may prove useful for investigating the dynamics of nanometre scale surface structures, where high beam coherence is necessary. Demonstrating single-shot diffraction is a significant step forward for cold atom electron sources, and supports the promise that they could complement solid state photocathode sources for use in ultrafast single-shot electron diffraction experiments.

\section*{Acknowledgements}
We thank L. J. Allen for helpful discussions and advice. BMS gratefully acknowledges the support of a University of Melbourne McKenzie Fellowship. This work was supported by the Australian Research Council Discovery Project DP140102102.

\section*{References}


\end{document}